\documentclass[
reprint,
superscriptaddress,
nofootinbib,
amsmath,amssymb,
aps,
longbibliography,
prb,
]{revtex4-2}

\usepackage{amsbsy,amssymb,amsmath,mathtools}
\usepackage{xcolor}
\usepackage{graphicx}
\usepackage{mathrsfs}

\usepackage{tikz}
\usetikzlibrary{decorations.pathmorphing}

\makeatletter
\PassOptionsToPackage{caption=false}{subfig} 
\usepackage{hyperref}
\hypersetup{
breaklinks=true,
colorlinks=true,
citecolor=blue,
linkcolor=black,
filecolor=black,
urlcolor=blue
}
\IfFileExists{lmodern.sty}{\usepackage{lmodern}}{}

\def\be{\begin{equation}}
\def\ee{\end{equation}}

\renewcommand{\bar}{\overline}

\renewcommand{\hat}{\widehat}

\usepackage{nicematrix}

\DeclareMathOperator{\Tr}{Tr}

\begin{document}

\title{Exact operator dynamics in Lindbladian Wess-Zumino-Witten conformal field theories}

\author{Qicheng Tang}
\affiliation{School of Physics, Georgia Institute of Technology, Atlanta, GA 30332, USA}

\author{Ruhanshi Barad}
\affiliation{School of Physics, Georgia Institute of Technology, Atlanta, GA 30332, USA}

\author{Xueda Wen}
\affiliation{School of Physics, Georgia Institute of Technology, Atlanta, GA 30332, USA}

\begin{abstract}

Understanding the time evolution of physical observables in open quantum many-body systems coupled to external environments is a natural and difficult problem, and exact results are still rare. In this work, we study this problem for Wess-Zumino-Witten (WZW) conformal field theories with Lindblad jump operators linear in Kac-Moody current modes. 
We investigate the exact operator dynamics generated by these Lindbladians, identifying classes of current operators whose Heisenberg equations close and can therefore be solved analytically using the underlying current algebra.
In Abelian $U(1)_k$ WZW theories, this closure of operator dynamics holds for arbitrary settings of jump rates and includes exactly tractable cooling dynamics. 
In contrast, for non-Abelian WZW theories, exact closure occurs only for symmetric current-mode dissipation, where upward and downward current-mode transitions occur with equal rates, and even then it leads to a simple closed evolution only for a single current operator.
Generic imbalances, including those needed for cooling, produce additional non-Abelian terms and prevent closure of the opeartor dynamics.
Consequently, the current algebra gives rise to a broad family of exactly solvable dissipative dynamics in the Abelian setting, whereas in the non-Abelian case it singles out only a special exactly solvable dynamics corresponding to an infinite-temperature bath.

\end{abstract}
\maketitle

\tableofcontents

\section{Introduction}

The dynamics of quantum many-body systems is generically influenced by coupling to external degrees of freedom. In many physically relevant settings, ranging from engineered quantum platforms such as cold-atom and quantum-optical systems \cite{muller2012engineered,reiter2017dissipative,Tomadin_2012} to mesoscopic transport in quantum materials, and even to certain semiclassical treatments of quantum fields in de Sitter space \cite{2025Takanori_Lindblad_holography}, this influence can often be modeled within the framework of a Lindblad master equation \cite{Breuer_2007, lindblad1976generators}. Although the Lindblad formalism is the standard language for describing Markovian open-system dynamics, exact analytic control over interacting many-body Lindbladians remains rare. The broadest systematic class of exactly solvable examples is provided by quasi-free fermionic and bosonic systems, where covariance-matrix methods and third quantization reduce the problem to effectively single-particle data \cite{prosen2008_thrid, prosen2010_XY, zhang2022_quadratic_lindblad, 2023Yamanaka}. Important interacting examples do exist, including boundary-driven integrable spin chains and other algebraically constrained constructions, but these remain comparatively sparse and model-dependent \cite{Znidaric_2010, Znidaric_2011, Medvedyeva_2016, Ilievski_2017, de_Leeuw_2021, Yamamoto_2022, Alba_2025}. It is therefore natural to ask whether there are further algebraic settings in which open quantum many-body dynamics can be treated exactly.

In parallel, conformal field theory (CFT) has long provided one of the most powerful frameworks for the precise description of the low-energy behavior of strongly correlated quantum matter at criticality.
Owing to the constraints of conformal symmetry, certain classes of non-equilibrium dynamics remain analytically tractable, even in non-unitary settings \cite{2023_Lotkov,
2024_Mao,2024_Su,
2024_Wen_exact,2025_Lapierre,2025_Lin,2026_Mao,2025_Ruhanshi,2026_Wang_Karch}.
Despite these developments, much less is known about dissipative dynamics described by Lindblad master equations, where the interplay between conformal symmetry and environmental coupling remains largely unexplored.

\smallskip

Motivated by the above question, in this work we introduce and study a family of Lindbladian dynamics built from the current algebra of Wess-Zumino-Witten (WZW) conformal field theories. 
It is noted that WZW CFTs arise in a wide variety of contexts, including critical spin chains, non-Abelian topological phases, and they furnish canonical examples of interacting conformal fixed points~\cite{Affleck_Haldane_1987_spin_wzw, Affleck_1986_realize_wzw, Witten_1988_Jones_Polynomial, MooreRead1991, 1999_Ardonne_Schoutens_spin_singlet, 2001_Ardonne_spin_singlet, Nielsen2011_spin_su2k, 2012_Thomale_spin_su2k, Tu_2012_project_spin_wzw, Tu_2014_spin_wzw, 2013_michaud_spin_wzw, 2017_Tsvelik_spin_CSL, Quella2019_spin_liquid, 2024_Simon_parton_wave_func, Liu_2025_wzw_CSL}.
By exploiting the algebraic structure of the Kac-Moody symmetry underlying WZW CFTs, we identify situations in which the operator dynamics can be solved exactly, thereby extending the notion of exact solvability to a broad class of dissipative conformal dynamics.

The starting point of this work is the idea behind third quantization. In quasi-free fermionic or bosonic Lindblad systems, a quadratic Hamiltonian together with jump operators linear in the microscopic creation and annihilation operators gives rise to a Liouvillian whose full spectrum can be determined exactly \cite{prosen2008_thrid, prosen2010_XY, prosen2010_spectral_lindblad, prosen2010_thirdquant_boson, zhang2022_quadratic_lindblad}. 
On the other hand, in a WZW CFT, through the Sugawara construction, the Hamiltonian can be expressed quadratically in the current modes \cite{francesco2012conformal}. This suggests that WZW models may provide a natural setting in which to revisit exact solvability in open quantum dynamics, with symmetry currents replacing canonical microscopic creation/annihilation operators. 
With this observation, we therefore consider Lindblad generators whose coherent part is given by the WZW Hamiltonian and whose jump operators are linear in the same current modes. We refer to the resulting generators as \textit{WZW current-mode Lindbladians}, or \textit{WZW current Lindbladians} for short. 

The aim of this paper is twofold. First, we formulate the general structure of WZW current-mode Lindbladians and determine in what sense exact solvability survives in this setting. In general, these Lindbladians do not admit the same kind of solution as in third quantization, and we do not obtain a general solution of the full Liouvillian spectrum. What remains exactly tractable, however, is the Heisenberg evolution of current modes and of operators built from them whenever the corresponding equations of motion close. In such cases, the time evolution can be derived exactly from the Kac-Moody current algebra. 
The notion of solvability established here is therefore not a full spectral solution of the Liouvillian, but rather an exact solution of the Heisenberg equations for (composite) operators constructed from current modes.


Second, we determine how far this exact solvability extends in concrete WZW theories and what dissipative behavior and steady-state information can still be extracted within this framework. In the Abelian $U(1)$ case, closure persists for arbitrary linear choices of current-mode jump operators, with cooling dynamics appearing as one natural and practically relevant solvable example. Since the $U(1)_k$ WZW model is equivalent to the compact free-boson CFT, this result is consistent with the general structure of solvable free-boson Lindbladians \cite{McDonald_2023},
while also demonstrating that the present framework encompasses a broad class of exactly tractable dissipative dynamics.
Moreover, we show that the equations of motion close not only for the current modes themselves but also for most composite operators built from multiple current modes, so that their time evolution can be determined exactly from the current algebra without any restriction on the initial state. Only a limited subclass of composite operators fails to admit such a direct algebraic treatment; in those cases, additional input, such as assumptions on the initial state, is required to simplify the dynamics further. 
The genuinely restrictive result arises in the non-Abelian case. For theories such as $SU(2)_k$ WZW models, we show that exact solvability survives only for a special symmetric choice in which the raising and lowering current modes enter the jump operators with equal weight. The equations of motion close only in the symmetric case, because only then do the extra terms produced by the non-Abelian commutation relations cancel. 
Any imbalance between the raising and lowering contributions spoils this cancellation, and therefore breaks exact solvability in the sense defined above. Thus, unlike in the Abelian case, an exactly solvable cooling setup cannot be obtained simply by tuning the jump rates of the current modes. The only evident exactly solvable case within this non-Abelian construction is instead the symmetric infinite-temperature generator. Consequently, although the Abelian theory admits exactly tractable cooling dynamics, the present framework does not yield an analogous exactly tractable protocol for non-Abelian ground-state preparation.

The remainder of this paper is organized as follows.
In Sec.~\ref{Sec:WZW_intro}, we review the basic structure and notation of Wess-Zumino-Witten (WZW) conformal field theories.
In Sec.~\ref{Sec:Lindblad_intro}, we introduce the Lindblad formalism and the notion of exact solvability used in this work, and in Sec.~\ref{Sec:Lindblad_WZW} we formulate the class of WZW current-mode Lindbladians. We then study exact operator dynamics in $U(1)_k$ WZW current-mode Lindbladians in Sec.~\ref{Sec:Lindblad_U(1)}, followed by the non-Abelian case of $SU(2)_k$ WZW current-mode Lindbladians in Sec.\ref{Sec:Lindblad_SU(2)}. In Sec.\ref{Sec:Lindblad_general}, we extend the analysis to exact single-current dynamics in general non-Abelian WZW current-mode Lindbladians. Finally, in Sec.~\ref{Sec:Discuss}, we conclude with a discussion of the implications of our results and several future directions.

\section{Wess-Zumino-Witten models and current algebra}
\label{Sec:WZW_intro}

WZW models are $1+1$-dimensional conformal field theories associated with a Lie group $G$ and an integer level $k$. Their defining feature is the presence of conserved currents valued in the Lie algebra $\mathfrak g=\operatorname{Lie}(G)$, which generate an affine Kac-Moody symmetry \cite{francesco2012conformal}. WZW models are both mathematically rich and physically important. They arise, for example, as effective descriptions of critical one-dimensional systems and as chiral edge theories of certain $2+1$-dimensional topological phases~\cite{Affleck_Haldane_1987_spin_wzw, Affleck_1986_realize_wzw, Witten_1988_Jones_Polynomial, MooreRead1991, 1999_Ardonne_Schoutens_spin_singlet, 2001_Ardonne_spin_singlet, Nielsen2011_spin_su2k, 2012_Thomale_spin_su2k, Tu_2012_project_spin_wzw, Tu_2014_spin_wzw, 2013_michaud_spin_wzw, 2017_Tsvelik_spin_CSL, Quella2019_spin_liquid, 2024_Simon_parton_wave_func, Liu_2025_wzw_CSL}. For the purposes of the present work, the key point is that the Hamiltonian is completely determined by the current operators and, through the Sugawara construction, is quadratic in these currents. 
This makes WZW models a natural setting in which to search for an interacting analogue of the quadratic structures familiar from third quantization \cite{prosen2008_thrid,prosen2010_thirdquant_boson}.

For simplicity, we present the formulas below for a single simple Lie algebra $\mathfrak g$.  Placing the theory on a circle of circumference $L$, the currents split into left- and right-moving sectors. The left-moving currents admit the mode expansion
\begin{equation}
	J^a(x,t) = \frac{2\pi}{L} \sum_{n\in\mathbb Z}
	J_n^a\, e^{-\frac{2\pi i n}{L}(x-vt)} \,,
\end{equation}
while the right-moving currents $\bar J^a(x,t)$ are defined analogously, with $x-vt$ replaced by $x+vt$. Their Fourier modes generate two commuting copies of the affine Kac-Moody algebra,
\begin{equation}\label{eq:KM_algebra}
	[J_m^a,J_n^b] = i \sum_{c=1}^{\dim \mathfrak g} f^{abc}\, J_{m+n}^c
	+ km\delta^{ab}\delta_{m+n,0} \,,
\end{equation}
with the same relation for $\bar J_n^a$, and
\begin{equation}
	[J_m^a,\bar J_n^b]=0.
\end{equation}
Here $f^{abc}\in\mathbb R$ are the structure constants of $\mathfrak g$ in an orthonormal basis, and they are totally anti-symmetric in the indices $a,b,c$. The integer $k$ is the level of the affine algebra, and the term $km\delta^{ab}\delta_{m+n,0}$ is its central term.

The conformal structure of the theory is encoded in the stress tensor, which is constructed from the currents through the \textit{Sugawara construction}. In mode form, the Virasoro generators are 
\begin{equation}
	L_m = \frac{1}{2(k+h^\vee)}
	\sum_{a=1}^{\dim\mathfrak g}
	\sum_{n\in\mathbb Z}
	: J_n^a J_{m-n}^a : \,,
\end{equation}
and similarly for $\bar L_m$, where $h^\vee$ is the dual Coxeter number of $\mathfrak g$, and $:\cdots:$ denotes normal ordering. In particular,
\begin{equation}
	L_0 = \frac{1}{2(k+h^\vee)}
	\sum_{a=1}^{\dim\mathfrak g}
	\sum_{n\in\mathbb Z}
	: J_n^a J_{-n}^a : \,,
\end{equation}
with an analogous expression for $\bar L_0$. The corresponding Virasoro central charges are
\begin{equation}
	c=\bar c=\frac{k\,\dim\mathfrak g}{k+h^\vee}.
\end{equation}
The Abelian $U(1)$ case is recovered by setting $f^{abc}=0$ and $h^\vee=0$ in the above definition. The algebra in \eqref{eq:KM_algebra} becomes $[J_m,J_n]=km\delta_{m+n,0}$, which is 
just a Heisenberg algebra.

The Hamiltonian of a WZW CFT on the cylinder is therefore
\begin{equation}\label{eq:wzw_ham}
	H_{\rm WZW} = \frac{2\pi v}{L}
	\left( L_0+\bar L_0-\frac{c+\bar c}{24} \right)
	= H_{\rm WZW}^{\rm hol} + H_{\rm WZW}^{\rm anti-hol} \,,
\end{equation}
with holomorphic part
\begin{equation}
	H_{\rm WZW}^{\rm hol}
	= \frac{2\pi v}{L}\sum_{a=1}^{\dim \mathfrak g}\frac{1}{2(k+h^\vee)}
	\sum_{m\in\mathbb Z} :J_m^a J_{-m}^a:
	-\frac{2\pi v}{L}\frac{c}{24} \,,
\end{equation}
and anti-holomorphic part
\begin{equation}
	H_{\rm WZW}^{\rm anti-hol}
	= \frac{2\pi v}{L}\sum_{a=1}^{\dim \mathfrak g}\frac{1}{2(k+h^\vee)}
	\sum_{m\in\mathbb Z} :\bar J_m^a \bar J_{-m}^a:
	-\frac{2\pi v}{L}\frac{\bar c}{24} \,.
\end{equation}
Thus the WZW Hamiltonian is explicitly quadratic in the current modes.

A particularly important consequence of the Sugawara construction is the commutator
\begin{equation}
	[L_m,J_n^a]=-n\,J_{m+n}^a,
\end{equation}
and in particular
\begin{equation}\label{eq:L0_commutator}
	[L_0,J_n^a]=-n\,J_n^a.
\end{equation}
Thus, if a state is an eigenstate of $L_0$, then acting with $J_n^a$ produces, whenever nonzero, a state whose $L_0$ eigenvalue is shifted by $-n$. In this sense, modes with $n>0$ lower the $L_0$ eigenvalue, whereas modes with $n<0$ raise it. Since the Hamiltonian on the cylinder is proportional to $L_0+\bar L_0$, this ladder structure of the current modes will play a central role in the Lindbladian constructions analyzed in the following sections.

\section{Lindbladian dynamics and exact solvability in operator dynamics}
\label{Sec:Lindblad_intro}

The Lindblad formalism provides an effective description of irreversible quantum dynamics for a subsystem coupled to an environment \cite{Breuer_2007}.
The state of the subsystem is described by a density matrix $\rho(t)$, obtained after tracing out the environmental degrees of freedom. In the Markovian regime, where memory effects of the environment are neglected, the time evolution of $\rho(t)$ is generated by a one-parameter semigroup of completely positive and trace-preserving maps,
\begin{equation}
	\rho(t)=e^{t\mathcal{L}}\rho(0),
\end{equation}
where $\mathcal{L}$ is the Lindbladian, also called the Liouvillian. In practice, this description is typically justified under the weak-coupling, Born-Markov, and secular approximations \cite{lindblad1976generators, gorini_1976CPTP}. The corresponding master equation takes the standard Gorini-Kossakowski-Lindblad-Sudarshan form
\begin{equation}
	\frac{d}{dt}\rho
	= \mathcal{L}[\rho]
	= -i[H,\rho] + \mathcal{D}[\rho] \,,
\end{equation}
with the dissipative term 
\begin{equation}
\label{Diss_term}
	\mathcal{D}[\rho] = 
	\sum_i \left(
	K_i \rho K_i^\dagger
	- \frac{1}{2}\{K_i^\dagger K_i,\rho\}
	\right) \,,
\end{equation}
where $H$ is a Hermitian Hamiltonian describing the coherent part of the time evolution, and $\{K_i\}$ are jump operators encoding dissipative processes induced by the environment.

To understand why this framework becomes exactly solvable in certain cases, it is useful first to distinguish operators from superoperators. Operators such as $H$ or $K_i$ act on state vectors in the physical Hilbert space. By contrast, the Lindbladian $\mathcal{L}$ acts on the density matrix itself, and is therefore a \emph{superoperator}: a linear map whose input and output are operators. More generally, the set of linear operators on the system Hilbert space may itself be viewed as a vector space, often called \emph{Liouville space}. Once a basis is chosen, any operator may be represented by its matrix elements and then vectorized by stacking those matrix elements into a single column vector. In this way, Liouville space may be regarded as a doubled Hilbert space.

This Liouville-space viewpoint is the starting point of \emph{third quantization}. By itself, however, vectorization is only a change of representation: it rewrites the master equation as a linear evolution equation on operator space, but it does not make a generic Lindbladian exactly solvable. The quasi-free fermionic and bosonic cases are special. There, the Hamiltonian is quadratic in the microscopic creation and annihilation operators, and each jump operator is linear in them. Left and right multiplication by these microscopic ladder operators and their adjoints then define natural maps on Liouville space. Because the underlying ladder operators satisfy canonical commutation or anticommutation relations, suitable combinations of these left- and right-acting maps obey corresponding canonical relations as superoperators. The crucial point is that a quadratic Hamiltonian together with linear jump operators makes every contribution to the Lindbladian quadratic in these canonical superoperators. Indeed, the coherent term $-i[H,\rho]$ is the difference between left and right multiplication by a quadratic operator, and is therefore quadratic. Likewise, since each $K_i$ is linear, the term $K_i\rho K_i^\dagger$ is bilinear in left and right multiplication, while $K_i^\dagger K_i$ is quadratic, so the anticommutator term is quadratic as well. The full Lindbladian can therefore be rewritten as a quadratic form on the full vectorized operator space \cite{prosen2008_thrid,prosen2010_XY,prosen2010_spectral_lindblad,zhang2022_quadratic_lindblad}. This is the sense in which third quantization extends the usual idea of second quantization: the many-body Lindbladian is reduced to a quadratic problem controlled by a single-particle coefficient matrix. A linear canonical transformation then brings this matrix to diagonal form when it is diagonalizable, or more generally to Jordan form. This reduction lifts to a many-body normal or Jordan form of the full Lindbladian, yielding the full spectrum of $\mathcal{L}$, steady states, and asymptotic relaxation properties in the quasi-free setting. At the level of states, the same quasi-free structure implies that Gaussian states remain Gaussian under time evolution.

In the present work, however, we adopt a different but related viewpoint. For the WZW current-mode Lindbladians studied below, the current modes satisfy an affine Kac-Moody algebra rather than canonical commutation or anticommutation relations, and therefore do not in general lead to a canonical quadratic structure on the full Liouville space. We therefore turn to the Heisenberg picture and ask instead whether the equation of motion of the time-evolved operator is closed in itself. The Heisenberg evolution is generated by the superoperator $\mathcal{L}^\dagger$ defined through
\begin{equation}
	\mathrm{Tr}\, \big( O\,\mathcal{L}[\rho] \big)
	=
	\mathrm{Tr}\, \big( \mathcal{L}^\dagger[O]\,\rho \big),
\end{equation}
so that an operator $O$ evolves as
\begin{equation}
	\frac{d}{dt}O
	=
	\mathcal{L}^\dagger[O]
	=
	i[H,O]
	+
	\mathcal{D}^\dagger[O],
\end{equation}
where
\begin{equation}
	\mathcal{D}^\dagger[O] = \sum_i\left(
	K_i^\dagger O K_i
	-\frac{1}{2}\{K_i^\dagger K_i,O\}
	\right).
\end{equation}

The notion of exact solvability used in this work is the following. For the WZW current-mode Lindbladians studied below, we ask whether the Heisenberg equations close for current modes and for composite operators built from them. Whenever this happens, the time evolution of these operators can be determined exactly from the Kac-Moody current algebra, without any assumption on the initial state. This does not amount to a solution of the full Lindbladian on Liouville space: closed Heisenberg equations by themselves do not in general determine the complete spectrum of $\mathcal{L}$, all steady states, or all relaxation modes. The problem addressed in the following sections is therefore to identify those WZW current-mode Lindbladians for which this closure occurs, and to determine what dynamical and steady-state information can be extracted from it.

\section{Construction of WZW current-mode Lindbladians}
\label{Sec:Lindblad_WZW}

In this section, we formulate the class of WZW current-mode Lindbladians studied in this work. 
They are constructed by taking the WZW Hamiltonian $H_{\rm WZW}$ in Eq.~\eqref{eq:wzw_ham} as the coherent part of the Lindbladian and choosing jump operators in \eqref{Diss_term} linear in the affine current modes as follows,
\begin{equation}
	K_q^a = \sqrt{\gamma(q, a)}\, J_q^a \,, 
	\quad 
	\bar{K}_q^a = \sqrt{\bar{\gamma}(q, a)}\, \bar{J}_q^a \,,
\end{equation}
where $q \in \mathbb{Z}$, $a = 1, \cdots, \dim \mathfrak g$.  
Since $J_{-q}^a$ ($q>0$) create descendants above the vacuum, and $J_{q}^a$ ($q>0$) annihilate descendants, 
the jump operators $K_q^a$ ($q>0$) correspond to loss (or cooling) processes, while $K_{-q}^a$ ($q>0$) correspond to gain (or pumping) processes. 
$\gamma(q,a)$ and $\bar{\gamma}(q,a)$ are independent non-negative jump rates for the holomorphic and anti-holomorphic components, respectively.
They characterize the strength of the system's coupling to the environment in the corresponding affine-current channels. In particular, $\gamma(q,a)$ with $q>0$ correspond to the loss rates, while $\gamma(q,a)$ with $q<0$ correspond to the gain rates, and similarly for the anti-holomorphic parts.

Since $(J_q^a)^\dagger = J_{-q}^a$, one has
\begin{equation}
	(K_q^a)^\dagger = \sqrt{\gamma(q,a)}\, J_{-q}^a  \,,
	\quad 
	(\bar{K}_q^a)^\dagger = \sqrt{\bar{\gamma}(q,a)}\, \bar{J}_{-q}^a  \,.
\end{equation}
The resulting Lindblad master equation of the density matrix $\rho$ of the WZW theory is
\begin{equation}
\frac{d\rho}{dt} = \mathcal{L}[\rho] = -i [H_{\rm WZW}, \rho] + \mathcal{D}[\rho],
\end{equation}
where the dissipative term is given by
\begin{equation}
\begin{aligned}
\mathcal{D}[\rho] & = \sum_{a=1}^{\dim \mathfrak{g}} \sum_{q\in\mathbb{Z}} \gamma(q,a) \left( J_q^a \rho J_{-q}^a - \frac{1}{2} \{ J_{-q}^a J_q^a, \rho \} \right) 
\\ & \quad + (J \to \bar{J} , \gamma \to \bar{\gamma})
\end{aligned}
\end{equation}

To make the resulting quadratic structure precise, it is convenient to introduce, for any operator $A$, the left- and right-multiplication superoperators
\begin{equation}
	\mathbb{L}_A[\rho] = A \rho, \qquad \mathbb{R}_A[\rho] = \rho A .
\end{equation}
They satisfy
\begin{equation}
	\mathbb{L}_{AB} = \mathbb{L}_A \mathbb{L}_B, \qquad
	\mathbb{R}_{AB} = \mathbb{R}_B \mathbb{R}_A .
\end{equation}
The Sugawara Hamiltonian in \eqref{eq:wzw_ham} is a sum of normal-ordered bilinears in the current modes, together with a central term, which is dropped out from the commutator, since it is a constant. The coherent contribution is then
\begin{equation}
	-i[H_{\rm WZW},\rho]
	=
	-i\bigl(\mathbb{L}_{H_{\rm WZW}}-\mathbb{R}_{H_{\rm WZW}}\bigr)[\rho]
\end{equation}
is quadratic in the multiplication superoperators built from the current modes. Likewise, because each jump operator $K_q^a$ is linear in a single current mode, every dissipative contribution is also quadratic, as
\begin{equation}
	\begin{aligned}
		&\quad \gamma(q,a) \left(
		J_{q}^a \rho J_{-q}^a
		-
		\frac{1}{2}\{J_{-q}^a J_q^a,\rho\}
		\right)
		\\ & = \gamma(q,a)
		\left(
		\mathbb{L}_{J_{q}^a}\mathbb{R}_{J_{-q}^a}
		-
		\frac{1}{2}\mathbb{L}_{J_{-q}^a}\mathbb{L}_{J_q^a}
		-
		\frac{1}{2}\mathbb{R}_{J_q^a}\mathbb{R}_{J_{-q}^a}
		\right)[\rho] \,,
	\end{aligned}
\end{equation}
so is the anti-holomorphic component. 
Hence the full Lindbladian $\mathcal{L}$ is quadratic in the left- and right-multiplication superoperators associated with the current modes. This is the precise sense in which a quadratic Hamiltonian together with jump operators linear in the current modes gives rise to a quadratic Lindbladian. The crucial difference from the standard quasi-free setting is that the current modes satisfy an affine Kac-Moody algebra rather than canonical commutation or anticommutation relations. In the Abelian case, the nonzero current modes can be rescaled to canonical bosonic oscillators by $J_m \to J_m / \sqrt{k}$, where $k$ is the level of affine algebra introduced in \eqref{eq:KM_algebra}. In the non-Abelian case, however, the structure-constant term in \eqref{eq:KM_algebra} obstructs such a reduction. Therefore, although the Lindbladian is quadratic in current-mode superoperators, it does not in general have the canonical quadratic form underlying standard third quantization for quasi-free fermionic or bosonic systems \cite{prosen2008_thrid}, nor does it provide a reduction to normal form for diagonalizing the Lindbladian.

As we have mentioned earlier, in this work, we adopt a different notion of exact solvability. Instead of solving the full spectrum of the Lindbladian, we seek for a closed Heisenberg equation for exactly solving the operator dynamics. In particular, under the WZW current Lindbladian constructed above, the Heisenberg evolution of a given operator $O$ is
\begin{equation}\label{eq:operator_Lind_WZW}
	\frac{dO}{dt} = \mathcal{L}^\dagger[O]
	= i [H_{\rm WZW}, O] + \mathcal{D}_{\rm WZW}^\dagger[O] ,
\end{equation}
Here $\mathcal{D}^\dagger$ represents the dissipative term in the Heisenberg equation of the operator evolution, and decomposes into holomorphic and anti-holomorphic contributions,
\begin{equation}
	\mathcal{D}_{\rm WZW}^\dagger[O]
	=
	\mathcal{D}_{\rm WZW, hol}^\dagger[O]
	+
	\mathcal{D}_{\rm WZW, anti-hol}^\dagger[O] ,
\end{equation}
with
\begin{equation}\label{eq:dissipator_commutator}
	\begin{aligned}
		\mathcal{D}_{\rm WZW, hol}^\dagger[O]
		&=
		\sum_{a=1}^{\dim \mathfrak{g}} \sum_{q \in \mathbb{Z}} \gamma(q,a)
		\left(
		J_{-q}^a O J_q^a
		-
		\frac{1}{2} \{ J_{-q}^a J_q^a, O \}
		\right)
		\\
		&=
		\sum_{a=1}^{\dim \mathfrak{g}} \sum_{q \in \mathbb{Z}} \frac{\gamma(q,a)}{2}
		\left(
		J_{-q}^a [O, J_q^a] + [J_{-q}^a, O] J_q^a
		\right) ,
	\end{aligned}
\end{equation}
and
\begin{equation}
	\mathcal{D}_{\rm WZW, anti-hol}^\dagger[O]
	=
	\mathcal{D}_{\rm WZW, hol}^\dagger[O]\bigl(J \to \bar J\bigr) .
\end{equation}
The second line in Eq.~\eqref{eq:dissipator_commutator} is simply a rearrangement of the dissipator into commutator form.

Since the holomorphic and anti-holomorphic current algebras commute, we may restrict attention to the holomorphic sector when studying chiral holomorphic operators \footnote{For a general local operator 
carrying both holomorphic and anti-holomorphic quantum numbers, the Lindbladian acts simultaneously through the 
$J_q^a$ and $\bar J_q^a$ sectors. Although the two current algebras commute, the resulting equations of motion for non-chiral operators generally involve both sectors and are therefore coupled.
}. 
For such an operator $O$, one has
\begin{equation}
	[\bar J_q^a,O]=0
	\qquad
	\text{for all } q\in\mathbb Z,\; a=1,\dots,\dim\mathfrak g,
\end{equation}
and therefore
\begin{equation}
	\mathcal{D}_{\rm WZW, anti-hol}^\dagger[O]=0.
\end{equation}
Moreover, the anti-holomorphic Hamiltonian $H_{\rm WZW}^{\rm anti-hol}$ also commutes with $O$. Hence, for the purpose of solving the dynamics of chiral holomorphic operators, it is sufficient to work with the reduced Heisenberg equation
\begin{equation}
	\frac{dO}{dt}
	=
	i[H_{\rm WZW}^{\rm hol},O]
	+
	\mathcal{D}_{\rm WZW, hol}^\dagger[O] .
\end{equation}
The anti-holomorphic sector is treated in exactly the same way after the replacement $J\to \bar J$. For simplicity, in the subsequent discussion of the operator dynamics, unless otherwise specified, we use $H_{\rm WZW}$ and $\mathcal{D}_{\rm WZW}$ to denote the holomorphic components of the full WZW Hamiltonian and the current-mode dissipator, respectively, and omit the subscript ``$\text{hol}$''.

As emphasized earlier, the notion of exact solvability used in this work does not refer to solving the full spectrum of the Lindbladian. Instead, we ask whether the Heisenberg evolution of a given operator closes within a finite-dimensional operator subspace. 
More precisely, given a set of operators $\{O_i\}_{i=1}^N$, we say that the dynamics closes if the adjoint Lindbladian maps this set back into its \textit{linear} span, 
\be
\mathcal L^\dag[O_i]=\sum_{j=1}^N M_{ij}\, O_j,
\ee
for some finite-dimensional matrix $M$.
In this case, the Heisenberg equations reduce to a closed system of linear ordinary differential equations.
The operator dynamics can therefore be determined exactly without requiring knowledge of the full Lindbladian spectrum.

Since we seek dynamics that are solvable purely within the affine current algebra, the natural class of solvable operators $O$ consists of those built from the current modes ${J_p^a}$, with $p \in \mathbb{Z}$ and $a = 1, \cdots, \dim \mathfrak{g}$. More generally, if the Lindbladian maps $O$, possibly a composite operator built from products of current modes, to a finite linear combination of (products of) current operators, the dynamics could reduce to a closed finite-dimensional system. In the next section, we illustrate the basic mechanism of exactly solvable operator dynamics using the $U(1)_k$ WZW CFT.

\section{Exact operator dynamics in $U(1)_k$ WZW current-mode Lindbladians}
\label{Sec:Lindblad_U(1)}

Having formulated the general construction of WZW current-mode Lindbladians and their quadratic structure, we now turn to the simplest concrete example, the $U(1)_k$ WZW model. Because the current algebra is Abelian, this case provides the most transparent setting to illustrate our notion of exact solvability for operator dynamics in WZW current Lindbladians.

For the $U(1)_k$ theory, the affine Kac-Moody algebra reduces to the Heisenberg algebra
\begin{align}
\label{eq:heisenberg_algebra}
	[J_m, J_n] = k\, m\, \delta_{m+n,0} \,, \quad
	[\bar J_m, \bar J_n] = k\, m\, \delta_{m+n,0},
\end{align}
while the two chiral sectors commute, i.e.,
$[\bar J_m, J_n] = 0$.
For a single chiral sector, this is simply the oscillator algebra of a free boson, up to the rescaling $J_m \to J_m/\sqrt{k}$. The $U(1)_k$ model therefore provides a natural starting point for extending the general discussion to products of several current modes. In this section, we first discuss the dynamics of a single current operator. We then study composite operators built from two or more current operators, and finally apply these results to a cooling protocol toward the critical ground state of the $U(1)_k$ WZW CFT.

\subsection{Lindbladian dynamics of single current operator}

Let us start with the evolution of a single current operator $O = J_p$ in $U(1)_k$ WZW Lindbladians. 
The coherent part is already closed on a single current mode. Indeed, as we have reviewed in Eq.~\eqref{eq:L0_commutator}, by the Sugawara construction,
\begin{equation}
	i[H_{{\rm WZW}, U(1)},J_p]
	=
	i\frac{2\pi v}{L}[L_0,J_p]
	=
	-\,i\frac{2\pi v}{L}\,p\,J_p .
\end{equation}
Thus any obstruction to closed single-current dynamics can only come from the dissipator.
Using the commutator form of the dissipator derived in Eq.~\eqref{eq:dissipator_commutator}, we obtain
\begin{equation}\label{eq:D_single_current_u1}
	\begin{aligned}
		\mathcal{D}^\dagger_{{\rm WZW}, U(1)}[J_p]
		&=
		\sum_{q\in\mathbb Z}
		\frac{\gamma(q)}{2}
		\left(
		J_{-q}[J_p,J_q]
		+
		[J_{-q},J_p]J_q
		\right)
		\\ &=
		\frac{kp}{2}\bigl[\gamma(-p)-\gamma(p)\bigr]J_p \,.
	\end{aligned}
\end{equation}
Combining the coherent and dissipative parts gives the closed equation of motion
\begin{equation}\label{eq:u1_jp_eom}
	\frac{dJ_p}{dt}
	=
	\mathcal L^\dagger[J_p]
	=
	\left[
	-i\frac{2\pi v}{L} + k\alpha_p 
	\right] pJ_p  \,,
\end{equation}
where we have defined
\begin{equation}
	\alpha_p
	=
	\frac{1}{2}
	\left[
	\gamma(-p)-\gamma(p)
	\right].
\end{equation}
Thus the mode-resolved solution is
\begin{equation}
\label{eq:U1_single_op_solution}
	J_p(t)
	= J_p(0)
	\exp\!\left[
	\left(
	-i\frac{2\pi v}{L}
	+
	k\alpha_p
	\right)p\, t
	\right] .
\end{equation}

Since the WZW CFT considered in this work is unitary, the level $k$ in $U(1)_k$ satisfies $k>0$. Then the single-current amplitude is bounded only if the real growth rate is non-positive, i.e.,
\begin{equation}
\label{Stable_condition}
	\alpha_p \,p
	=
	\frac{1}{2}
	\left[
	\gamma(-p)-\gamma(p)
	\right]p
	\leq 0 , \quad \forall \, p\in \mathbb Z.
\end{equation}
This condition will be referred to as the \emph{stability condition} \cite{prosen2010_thirdquant_boson,2023_Kim}.
Equivalently,
\begin{equation}
	\gamma(|p|)
	\ge
	\gamma(-|p|),
	\qquad p\neq 0 .
\end{equation}
If this condition is violated, the corresponding current mode grows exponentially in the Heisenberg picture. The system is therefore unstable: the environment injects excitations faster than it removes them, so no steady state can be reached. Physically, the stability condition requires that, for the system to reach steadiness, the dissipative removal of excitations from the current mode must at least balance their dissipative injection.

\subsection{Lindbladian dynamics of two current operators}

We next turn to products of two current operators. The key tool is the general product rule for the adjoint Lindbladian. For two operators $A$ and $B$,
\begin{equation}
	\frac{d(AB)}{dt}
	=
	\mathcal L^\dagger[AB]
	=
	i[H,AB]+\mathcal D^\dagger[AB] ,
\end{equation}
where
\begin{equation}
	\mathcal D^\dagger[AB]
	=
	\sum_i
	\left(
	K_i^\dagger AB K_i
	-\frac{1}{2}\{K_i^\dagger K_i,AB\}
	\right) .
\end{equation}
For the coherent part, one has
$[H,AB]=[H,A]B+A[H,B]$.
For the dissipative part, we have
\begin{equation}
	\mathcal D^\dagger[AB]
	=
	\mathcal D^\dagger[A]B
	+
	A\mathcal D^\dagger[B]
	+
	\sum_i [K_i^\dagger,A][B,K_i] .
\end{equation}
Therefore,
\begin{equation}
\label{eq:iteration_TwoOp_lind}
	\mathcal L^\dagger[AB]
	=
	\mathcal L^\dagger[A]B
	+
	A\mathcal L^\dagger[B]
	+
	\sum_i [K_i^\dagger,A][B,K_i] .
\end{equation}
This identity is not only useful for the dynamics of two operators, but also provides the recursive step for the multi-current analysis below.

For the $U(1)_k$ WZW model, we take $A=J_p$ and $B=J_r$. Using the single-current result above, we obtain
\begin{equation}\label{eq:u1_jpjr_TwoPoint_eom}
	\begin{aligned}
		& \quad \frac{d[J_pJ_r]}{dt}
		=
		\mathcal L^\dagger[J_pJ_r]
		\\ & =
		\mathcal L^\dagger[J_p]J_r
		+
		J_p\mathcal L^\dagger[J_r]
		+
		\sum_{q\in\mathbb Z}\gamma(q)[J_{-q},J_p][J_r,J_q]
		\\
		&=
        \sum_{\mathfrak{m}=p,r} \left( -i\frac{2\pi v}{L} + k \alpha_\mathfrak{m} \right) \mathfrak{m} J_pJ_r
        + 
		\gamma(p)k^2p^2\,\delta_{p,-r} \,.
	\end{aligned}
\end{equation}
This differential equation has the standard form
\begin{equation}
	\frac{dy}{dt}=cy+b ,
\end{equation}
with
\begin{equation}
	\begin{aligned}
		c(p,r)&=-i\frac{2\pi v}{L}(p+r)+(\alpha_p p+\alpha_r r)k,\\
		b(p,r)&=\gamma(p)k^2p^2\,\delta_{p,-r},
	\end{aligned}
\end{equation}
and initial condition $[J_pJ_r](t=0)=[J_pJ_r](0)$. Hence
\begin{equation}\label{eq:TwoOperator_u1k}
	[J_pJ_r](t)
	=
	-\frac{b(p,r)}{c(p,r)}
	+
	\left[
	[J_pJ_r](0)+\frac{b(p,r)}{c(p,r)}
	\right]e^{c(p,r)t}.
\end{equation}
Again, bounded evolution requires
\begin{equation}
	\alpha_p p+\alpha_r r\leq 0 .
\end{equation}
This condition is automatically satisfied whenever the stability condition Eq.~\eqref{Stable_condition} holds.
Nevertheless, the two-operator dynamics is exact solvable as a consequence of the closed equation of motion.

\subsection{Lindbladian dynamics of multiple current operators}

We now extend the discussion to ordered products of arbitrarily many current operators. The same product rule allows the equation of motion for an $N$-current operator to be built recursively from that of an $(N-1)$-current operator. Let
\begin{equation}
	A=J_{p_1}J_{p_2}\cdots J_{p_{N-1}},
	\qquad
	B=J_{p_N}.
\end{equation}
Iterating the relation in \eqref{eq:iteration_TwoOp_lind}, 
we obtain
\begin{equation}\label{eq:EoM_multi_u1k}
	\begin{aligned}
		& \quad \frac{d}{dt} \left[ \prod_{\ell = 1}^{N} J_{p_\ell} \right] 
		= \mathcal{L}^\dagger \left[ \prod_{\ell = 1}^{N} J_{p_\ell} \right]
		\\ &=
		\mathcal{L}^\dagger [A] B
		+
		A \mathcal{L}^\dagger[B]
		+
		\sum_{q \in \mathbb{Z}} \gamma(q) [J_{-q}, A] [B, J_q]
		\\
		&=
		\sum_{\mathfrak{m}=1}^{N}
		\left(
		-i \frac{2\pi v}{L}
		+
		\alpha_{p_{\mathfrak{m}}} k
		\right)
		p_{\mathfrak{m}}
		\left( \prod_{\ell=1}^{N} J_{p_\ell} \right)
		\\
		&\quad
		-
		k^2
		\sum_{1 \le i < j \le N}
		p_i p_j \,\gamma(p_i)\,\delta_{p_i,-p_j}
		\prod_{\substack{\ell=1 \\ \ell \neq i,j}}^{N,\mathrm{ord}} J_{p_\ell} .
	\end{aligned}
\end{equation}
Here the superscript ``ord'' means that the product is taken in the original operator order. More precisely, $\prod_{\ell \neq i,j}^{N,\mathrm{ord}} J_{p_\ell}$ denotes the ordered sequence $J_{p_1}\cdots J_{p_N}$ with $J_{p_i}$ and $J_{p_j}$ omitted, while preserving the relative order of all remaining factors. 

Equation~\eqref{eq:EoM_multi_u1k} shows that the multi-current dynamics has a recursive structure. The first term is linear in the full ordered product, while the second term comes from the central term of the Kac-Moody algebra and contributes only when opposite-mode pairs $J_p$ and $J_{-p}$ are present. 
Therefore, whenever no such pairs occur, the equation closes directly on the ordered product itself and is solved by
\begin{equation}
	\left[ \prod_{\ell = 1}^{N} J_{p_\ell} \right] (t)
	=
	e^{
	\sum_{\mathfrak{m}=1}^{N}
	\left(
	-i \frac{2\pi v}{L}
	+
	\alpha_{p_{\mathfrak{m}}} k
	\right)
	p_{\mathfrak{m}} t
	}
	\left[ \prod_{\ell = 1}^{N} J_{p_\ell} \right] (t=0) \,.
\end{equation}

When such pairs occur, the equation of motion is no longer closed only in the evolved operator, but involves additional terms of operators with less current modes. This makes the solution more hard to be analytically accessed. 
Let us adopt the case of $N=3$, i.e. the evolved operator $O=J_{p_1}J_{p_2}J_{p_3}$ is a product of three current operators, as a concrete example. In this case,
\begin{equation}
\begin{aligned}
& \quad \frac{d O}{dt} 
= \mathcal{L}^\dagger[ J_{p_1} J_{p_2} J_{p_3} ] 
\\ & = \sum_{\mathfrak{m}=1}^{3}
\left(
-i \frac{2\pi v}{L}
+
\alpha_{p_{\mathfrak{m}}} k
\right)
p_{\mathfrak{m}}
O
+ k^2 p_1^2 \gamma(p_1) \delta_{p_1, -p_2} J_{p_3} 
\\ & \quad + k^2 p_1^2 \gamma(p_1) \delta_{p_1, -p_3} J_{p_2}
+ k^2 p_2^2 \gamma(p_2) \delta_{p_2, -p_3} J_{p_1}.
\end{aligned}
\end{equation}
Recall that we have solved the single-operator dynamics in Eq.~\eqref{eq:U1_single_op_solution}, giving 
\begin{equation}
	\frac{d O}{dt}  
	- c O 
	= \sum_{\mathfrak{m}=1}^{3} f_\mathfrak{m} J_{p_\mathfrak{m}} 
	= \sum_{\mathfrak{m}=1}^{3} f_\mathfrak{m} J_{p_\mathfrak{m}}(0) e^{ g_{\mathfrak{m}} t } \,,
\end{equation}
where
\begin{equation}
	c = \sum_{\mathfrak{m}=1}^{3} g_{\mathfrak{m}}  \,, 
	\quad
	g_{\mathfrak{m}} 
	= \left( -i \frac{2\pi v}{L} + \alpha_{p_{\mathfrak{m}}} k \right) p_{\mathfrak{m}} \,, 
\end{equation}
and
\begin{equation}
f_{\mathfrak{m}} = \begin{cases}
    k^2 p_2^2 \gamma(p_2) \delta_{p_2, -p_3} \,, & \mathfrak{m} = 1 \\ 
    k^2 p_1^2 \gamma(p_1) \delta_{p_1, -p_3} \,, & \mathfrak{m} = 2 \\ 
    k^2 p_1^2 \gamma(p_1) \delta_{p_1, -p_2} \,, & \mathfrak{m} = 3
\end{cases}
\end{equation}
The solution is given by
\begin{equation}
	O(t) = e^{ct} \, O(0) + \sum_{\mathfrak{m}=1}^{3} f_{\mathfrak{m}} J_{p_{\mathfrak{m}}}(0) \frac{e^{g_\mathfrak{m} t} - e^{ct}}{g_\mathfrak{m} - c} \,,
\end{equation}
where $O = J_{p_1} J_{p_2} J_{p_3}$. 
In principle, the explicit solution for a general composite operator built from $N$ current modes can be obtained recursively from the solution for operators with $N-2$ current modes, ultimately reducing the problem to the one-operator and two-operator cases, although carrying out this procedure for arbitrary $N$ may be not easy in practice.

A particularly simple corollary arises for Gaussian initial states. In that case, Wick’s theorem reduces all expectation values to one- and two-point functions, which have already been solved exactly in the previous subsections. The expectation-value dynamics in the Gaussian case is therefore completely determined by the single-current and two-current sectors.

\subsection{Cooling to the critical ground state and mixing time}
\label{Sec:Cooling}

We now use the exact operator dynamics of the $U(1)_k$ WZW Lindbladian to analyze a dissipative cooling protocol toward the critical ground state. The jump operators are chosen to be ladder operators that induce unidirectional transitions from higher-energy levels to lower-energy levels, thereby driving the system toward its ground state. In the present setting, this is implemented by taking the positive current modes $J_q$ with $q>0$ as jump operators, so that the cooling dissipator is
\begin{equation}
	\mathcal{D}_{ U(1), {\rm cool}}^\dagger [O]
	=
	\sum_{q>0} \gamma(q)
	\left(
	J_{-q} O J_q
	-\frac{1}{2}\{J_{-q}J_q,O\}
	\right),
\end{equation}
where $\gamma(q)>0$ characterize the cooling rate. Since the zero mode $J_0$ does not play any role in the cooling dynamics discussed below, we omit it for simplicity.

Because $J_0$ is conserved, the zero-mode charge is preserved by the evolution. The critical ground state of the $U(1)_k$ WZW model has zero $J_0$ charge, so only initial states satisfying the same zero-mode condition can relax to it under this dynamics.

For the cooling dissipator above, Eq.~\eqref{eq:TwoOperator_u1k} gives
\begin{equation}
\label{eq:JpJr}
	\begin{aligned}
		[J_pJ_r](t)
		&=
		-\frac{\gamma(p)k^2p^2\delta_{p,-r}}
		{-i\frac{2\pi v}{L}(p+r)-\frac{k}{2}\left(\gamma(|p|)|p|+\gamma(|r|)|r|\right)}
		\\
		&\quad
		+
		e^{-i\frac{2\pi v}{L}(p+r)t-\frac{k}{2}\left(\gamma(|p|)|p|+\gamma(|r|)|r|\right)t}
		\Bigg[
		[J_pJ_r](0)
		\\ & \qquad +
		\frac{\gamma(p)k^2p^2\delta_{p,-r}}
		{-i\frac{2\pi v}{L}(p+r)-\frac{k}{2}\left(\gamma(|p|)|p|+\gamma(|r|)|r|\right)}
		\Bigg] .
	\end{aligned}
\end{equation}

In particular, for $p=-r$ this simplifies to
\begin{equation}
	[J_p J_{-p}](t)
	=
	\begin{cases}
		kp + e^{-kp\gamma(p)t}\left([J_pJ_{-p}](0)-kp\right), & p>0,\\[4pt]
		e^{-k|p|\gamma(|p|)t}\,[J_pJ_{-p}](0), & p<0 ,
	\end{cases}
\end{equation}
where we used $\gamma(p)=0$ for $p<0$. 
Thus, for $p>0$ the expectation value of operator $J_pJ_{-p}$ relaxes exponentially to the steady value $kp$, while for $p<0$ it decays exponentially to zero. Moreover, for $p \neq -r$, Eq.~\eqref{eq:JpJr} indicates an exponential decay to zero. In summary, at the late time $t \to +\infty$, the two-point correlators reach steady values as follows,
\be
\langle J_p J_r\rangle(t\to\infty) = \begin{cases}
    kp , & p=-r>0 \\[4pt]
    0, & p\neq -r \,\text{ or }\, p=-r<0 \\[4pt]
    \langle J_0(0)\rangle ^2 , & p=r=0
\end{cases}
\ee
When the initial state is in the zero-charge sector, i.e. $\langle J_0(0)^m \rangle = 0$ for positive integer $m$, the late-time two-point correlators have the identical expectation value as the critical ground state of the $U(1)_k$ WZW model. Below we will show that the Lindbladian dynamics indeed cools to the critical ground state in a logarithmic time by investigating its trace distance with the time-evolving state $\rho(t)$.

To bound the cooling time, 
or equivalently the mixing time that characterizes the time scale to approach the steady state (the ground state in the present setting),
we consider the particle-number operator
\begin{equation}
	\hat N
	=
	\sum_{n=1}^{\infty}\frac{1}{kn}J_{-n}J_n .
\end{equation}
If $|G\rangle$ denotes the target ground state, then Markov's inequality gives
\begin{equation}
	\begin{aligned}
		1-\langle G|\rho(t)|G\rangle
		\leq &
		\Tr[\hat N\rho(t)]
		=
		\sum_{n=1}^{\infty}\frac{1}{kn}\Tr\!\left([J_{-n}J_n](t)\rho(0)\right)
		\\
		&=
		\sum_{n=1}^{\infty}\frac{1}{kn}
		\Tr\!\left(e^{-kn\gamma(n)t}[J_{-n}J_n](0)\rho(0)\right)
		\\
		&\leq
		\Tr[\hat N\rho(0)]\,e^{-k\min_{n\geq 1}[n\gamma(n)]\,t}.
	\end{aligned}
\end{equation}
Defining
\begin{equation}
	\Delta
	=
	k\,\min_{n\geq 1}[n\gamma(n)],
\end{equation}
we see that $\Delta$ is the slowest decay rate among the current modes. In the special case $\gamma(n)=\gamma$ for all $n>0$, this reduces to
\begin{equation}
	\Delta = k\gamma .
\end{equation}

Next, by the Fuchs--van de Graaf inequality, the trace distance to the target ground state obeys
\begin{equation}
	\begin{aligned}
		D(\rho(t),|G\rangle\langle G|)
		\leq&
		\sqrt{1-\langle G|\rho(t)|G\rangle}
		\\ &\leq
		\sqrt{\Tr[\hat N\rho(0)]}\,e^{-\frac{1}{2}\Delta t}.
	\end{aligned}
\end{equation}
Therefore, for an initial state with the same conserved $J_0$ charge as the target ground state and finite $\Tr[\hat N\rho(0)]$, the mixing time $\tau_{\rm mix}$ defined by $D(\rho(t),|G\rangle\langle G|)\leq \eta$ is upper bounded by
\begin{equation}
	\tau_{\rm mix}
	\leq
	2\Delta^{-1}
	\log\!\left(
	\frac{\sqrt{\Tr[\hat N\rho(0)]}}{\eta}
	\right).
\end{equation}
This gives an explicit analytic bound on the mixing time toward the critical ground state in the present cooling scheme.

\smallskip
Before concluding this section, we emphasize that the flexibility in choosing the jump rates in the $U(1)_k$ WZW CFT allows one to engineer a variety of physically interesting steady states. For instance, by selecting jump rates that satisfy the detailed-balance condition, the Lindbladian dynamics can be designed to drive the system toward a thermal Gibbs state at finite temperature.

\section{Exact operator dynamics in $SU(2)_k$ WZW current-mode Lindbladians}
\label{Sec:Lindblad_SU(2)}

We now turn to the simplest non-Abelian example, the $SU(2)_k$ WZW current-mode Lindbladian. The holomorphic and anti-holomorphic currents generate two commuting copies of the affine $\widehat{\mathfrak{su}}(2)_k$ Kac--Moody algebra,
\begin{equation}\label{eq:su2_KM}
	\begin{aligned}
		[J_m^a, J_n^b] &= i \epsilon^{ab}{}_c J_{m+n}^c + \frac{k\,m}{2}\,\delta^{ab}\,\delta_{m+n,0} \,, \\
		[\bar J_m^a, \bar J_n^b] &= i \epsilon^{ab}{}_c \bar J_{m+n}^c + \frac{k\,m}{2}\,\delta^{ab}\,\delta_{m+n,0} \,,
	\end{aligned}
\end{equation}
with
\begin{equation}
	[J_m^a,\bar J_n^b]=0 \,.
\end{equation}
Here $a,b,c\in\{1,2,3\}$ label the Lie-algebra components, and $\epsilon^{ab}{}_c$ are the structure constants of $\mathfrak{su}(2)$. The essential new feature, compared with the Abelian $U(1)_k$ case, is the structure-constant term $i\epsilon^{ab}{}_c J_{m+n}^c$. It is this term that encodes the non-Abelian Lie-algebra structure and, in the Lindbladian dynamics, generates additional products of current modes. As a result, closure of the Heisenberg equations becomes much more restrictive than in the Abelian theory.

The discussion below proceeds in two steps. We first analyze the evolution of a single current mode and show that, in the $SU(2)_k$ case, symmetry of the jump rates under $q\to -q$ is enough to make the equation of motion of each single current mode closes on itself. We then consider operators composed of two current modes and demonstrate that this closure is generically lost, even for the most symmetric choice of jump rates.  This already exhibits the basic obstruction in the non-Abelian case. As before, we restrict attention to the holomorphic sector; the anti-holomorphic analysis follows identically upon replacing $J$ with $\bar J$.

\subsection{Lindbladian dynamics of a single current operator}

We begin with a single holomorphic current mode $O=J_p^b$. 
The coherent part is already closed:
\begin{equation}
	i[H_{\rm WZW},J_p^b]
	=
	i\frac{2\pi v}{L}[L_0,J_p^b]
	=
	-\,i\frac{2\pi v}{L}\,p\,J_p^b .
\end{equation}
The nontrivial question is therefore whether the dissipator preserves this form.
Specializing the general Heisenberg-picture dissipator to the $\widehat{\mathfrak{su}}(2)_k$ algebra gives
\begin{equation}\label{eq:D_single_current_su2}
	\begin{aligned}
		&\quad
        \mathcal{D}_{{\rm WZW},SU(2)}^\dagger[J_p^b]
		\\ &=
		\sum_{a,c=1}^{3}\sum_{q\in\mathbb Z}
		\frac{i}{2}\epsilon^{abc}\gamma(q,a)
		\left(
		-\,J_{-q}^aJ_{p+q}^c
		+
		J_{p-q}^cJ_q^a
		\right)
		\\
		&\quad
		+
		\frac{kp}{4}\bigl[\gamma(-p,b)-\gamma(p,b)\bigr]J_p^b .
	\end{aligned}
\end{equation}
For generic rates $\gamma(q,a)$, the first term is bilinear in the current modes and the equation of motion of $J_p^b$ is therefore not closed on single current operators.

To remove these bilinear terms, we impose symmetry under $q\to -q$,
\begin{equation}
\label{Rate_su2}
	\gamma(-q,a)=\gamma(q,a)\ge 0,
\end{equation}
and define
\begin{equation}
	\Gamma(a)=\frac{1}{2}\gamma(0,a)+\sum_{q>0}\gamma(q,a).
\end{equation}
Pairing the contributions with $q$ and $-q$, the bilinear terms combine into commutators and reduce to
\begin{equation}
	\frac{1}{2}\sum_{q\in\mathbb Z}\gamma(q,a)
	\left(
	-\,J_{-q}^aJ_{p+q}^c
	+
	J_{p-q}^cJ_q^a
	\right)
	=
	-\,i\sum_{d=1}^{3}\epsilon^{acd}\Gamma(a)J_p^d .
\end{equation}
Substituting this into Eq.~\eqref{eq:D_single_current_su2}, and noting that the central term vanishes once $\gamma(-p,b)=\gamma(p,b)$, we obtain
\begin{equation}
	\mathcal{D}_{{\rm WZW},SU(2)}^\dagger[J_p^b]
	=
	\sum_{d=1}^{3}
	\left(
	\sum_{a,c=1}^{3}\epsilon^{abc}\epsilon^{acd}\Gamma(a)
	\right)J_p^d .
\end{equation}
For the $SU(2)$ case, the tensor contraction can be evaluated explicitly:
\begin{equation}
	\sum_{a,c=1}^{3}\epsilon^{abc}\epsilon^{acd}\Gamma(a)
	=
	\left[
	\Gamma(b)-\sum_{a=1}^{3}\Gamma(a)
	\right]\delta^{bd}.
\end{equation}
It follows that the dissipator is already diagonal in the Cartesian basis,
\begin{equation}
\label{eq:single_current_su2_general_rate}
	\mathcal{D}_{{\rm WZW},SU(2)}^\dagger[J_p^b]
	=
	-\lambda_b\,J_p^b ,
	\quad
	\lambda_b=\sum_{a=1}^{3}\Gamma(a)-\Gamma(b).
\end{equation}
Consequently, 
the full Heisenberg equation closes on each individual current mode:
\begin{equation}
	\frac{dJ_p^b}{dt}
	=
	-\left(
	i\frac{2\pi v}{L}p+\lambda_b
	\right)J_p^b.
\end{equation}
The resulting dynamics is therefore exactly solvable, with solution
\begin{equation}
	J_p^b(t)
	=
	\exp\!\left[
	-\left(
	i\frac{2\pi v}{L}p+\lambda_b
	\right)t
	\right]J_p^b(0).
\end{equation}

A further simplification occurs when the rates are also independent of the Lie-algebra index,
\begin{equation}
	\gamma(q,a)=\gamma(q)=\gamma(-q).
\end{equation}
Then $\Gamma(1)=\Gamma(2)=\Gamma(3)\equiv \Gamma$, so that
\begin{equation}
	\lambda_1=\lambda_2=\lambda_3=2\Gamma=\sum_{q\in\mathbb Z}\gamma(q),
\end{equation}
and Eq.~\eqref{eq:single_current_su2_general_rate} reduces to
\begin{equation}
\label{eq:single_operator_EoM_su2}
	\frac{dJ_p^b}{dt}
	=
	\mathcal{L}^\dagger[J_p^b]
	=
	-\left[
	i\frac{2\pi v}{L}p
	+
	\sum_{q\in\mathbb Z}\gamma(q)
	\right]J_p^b .
\end{equation}
Thus, unlike the Abelian case where closure of equation of motion is automatic, the non-Abelian $SU(2)_k$ WZW CFT requires the additional condition of symmetric jump rates. Under this condition, the equation of motion for each single current mode closes on itself, rendering the dynamics exactly solvable.

\subsubsection{Difference between Abelian and non-Abelian cases}
\label{Sec:Abelian_non-Abelian}

Although the single-current dynamics is exactly solvable in both the $U(1)_k$ and $SU(2)_k$ WZW theories when the jump rates satisfy the symmetry condition $\gamma(p)=\gamma(-p)$, the resulting dynamics is qualitatively different in the two cases. For the Abelian $U(1)_k$ theory, the symmetric choice of rates eliminates the dissipative contribution entirely, and the single-current evolution in \eqref{eq:U1_single_op_solution} remains purely coherent, exhibiting no damping. By contrast, in the non-Abelian $SU(2)_k$ theory, the same symmetry condition removes only the central contribution to the dissipator. The non-Abelian part of the current algebra continues to generate dissipative dynamics, so that the equation of motion remains closed but generally exhibits exponential damping. We now analyze the origin of this qualitative difference in detail.

The qualitative difference between the Abelian and non-Abelian cases originates from the structure of the current algebra. In the $U(1)_k$ case, the current algebra is purely central, as shown in \eqref{eq:heisenberg_algebra}.
As a result, the dissipative contribution to the evolution of a single current mode is entirely controlled by the imbalance between the rates
$\gamma(-p)$ and $\gamma(p)$.
When the rates are symmetric, i.e., $\gamma(-p)=\gamma(p)$, this contribution cancels exactly, and the dissipator acts trivially on $J_p$.
The single-current dynamics is therefore purely coherent, governed only by the Hamiltonian evolution.

\smallskip
By contrast, in the $SU(2)_k$ theory, the affine current algebra contains a genuinely non-Abelian contribution, as shown in \eqref{eq:su2_KM}.
The symmetry of the rates under $q\to -q$ in \eqref{Rate_su2} removes the contribution associated with the central term, but it does not eliminate the structure-constant term. Instead, the bilinear terms in the dissipator combine into commutators, whose non-Abelian part acts nontrivially on the Lie-algebra indices of the current.
This produces an effective depolarization in the adjoint representation in \eqref{eq:single_current_su2_general_rate}, 
which can be rewritten as
\be
\mathcal D^\dagger[J_p^b]=
-\lambda_b J_p^b,
\quad
\lambda_b=\sum_{a\neq b}\Gamma(a).
\ee
Therefore, $J_p^b$ decays unless the dissipator contains only jump operators in the same Lie-algebra direction $b$.
To illustrate this, consider the case in which jump operators are introduced only in the $b=3$ component, namely $\Gamma(1)=\Gamma(2)=0$ and $\Gamma(3)\equiv \Gamma> 0$. Then the equation of motion for $J_p^3$ becomes
\begin{equation}
	\frac{dJ_p^3}{dt}
	=
	-i \left(
	\frac{2\pi v}{L}p
	\right)J_p^3,
\end{equation}
which is purely coherent and exhibits no damping. In contrast,
\begin{equation}
	\frac{dJ_p^a}{dt}
	=
	- \left(i
	\frac{2\pi v}{L}p+\Gamma
	\right)J_p^a,\quad a=1, 2,
\end{equation}
so the transverse current components decay exponentially in time.
Therefore, for general choices of jump operators with symmetry $p\to -p$ in the jump rates, while the equation of motion remains closed on each single current mode, the corresponding amplitude generally decays. The damping is not caused by an imbalance between upward and downward modes, but by the noncommutativity of different $SU(2)$ current components. 
That is, dissipation in one Lie-algebra direction acts nontrivially on currents in other directions through the commutation relations, producing decay even in the symmetric-rate regime.
This mechanism is not specific to $SU(2)_k$
and persists in general non-Abelian WZW theories, as we will discuss in Sec.\ref{Sec:Lindblad_general}.

As a final remark, we note that the condition \eqref{Rate_su2}, together with its Abelian counterpart $\gamma(-p)=\gamma(p)$, may be viewed as the infinite-temperature limit of the detailed-balance condition. However, because the Hilbert space of a WZW CFT is infinite dimensional, symmetric rates do not in general imply relaxation to a thermal \textit{steady} state.
 Instead, the system can continue to absorb energy through balanced upward and downward transitions, preventing convergence to a normalizable thermal state.
Moreover, as discussed above, the resulting dynamics depends sensitively on the underlying current algebra: in the Abelian $U(1)_k$ theory the single-current modes remain undamped, whereas in the non-Abelian $SU(2)_k$ theory the structure constants generate additional depolarization and lead to exponential decay of current modes.

\subsection{Lindbladian dynamics of two current operators}

We next ask whether the exact solvability extends to products of two current modes. Since closure is most likely to occur for the most symmetric choice of jump rates, it suffices to consider
\begin{equation}
\gamma(q,a)=\gamma(q)=\gamma(-q)\ge 0.
\end{equation}
If the equation of motion fails to close even in this maximally symmetric setting, then closure cannot be restored by a more general choice of rates.

Consider
\begin{equation}
	O=J_p^bJ_r^c .
\end{equation}
Using the iteration identity \eqref{eq:iteration_TwoOp_lind} together with the single-current result \eqref{eq:single_operator_EoM_su2}, we obtain
\begin{equation}
	\begin{aligned}
		\mathcal{L}^\dagger[J_p^bJ_r^c]
		&=
		-\left[
		i\frac{2\pi v}{L}(p+r)
		+
		2\sum_{q\in\mathbb Z}\gamma(q)
		\right]J_p^bJ_r^c
		\\
		&\quad
		+
		\sum_{q\in\mathbb Z}\gamma(q)
		\left[
		\delta^{bc}\sum_{d=1}^{3}J_{p-q}^dJ_{r+q}^d
		-
		J_{p-q}^cJ_{r+q}^b
		\right]
		\\
		&\quad
		+
		\frac{k}{2}\bigl[r\gamma(r)-p\gamma(p)\bigr]
		\sum_{d=1}^{3} i\epsilon^{cbd}J_{p+r}^d
		\\
		&\quad
		+
		\frac{k^2p^2}{4}\,\gamma(p)\,\delta^{bc}\delta_{p,-r}.
	\end{aligned}
\end{equation}

The first line is the direct coherent evolution and damping of the original operator $J_p^bJ_r^c$. The second line is the genuinely non-Abelian contribution: it couples the original operator to infinitely many shifted products $J_{p-q}^dJ_{r+q}^e$. The third line mixes in a single current operator through the central extension, and the last line is a $c$-number term.

Therefore, even for the maximally symmetric choice of jump rates under which the single-current dynamics is exactly solvable, the equation of motion of a generic product $J_p^bJ_r^c$ is not closed on itself, nor on any finite family of two-current operators with fixed mode numbers. This is the first clear sign that exact solvability in the non-Abelian case is substantially more restricted beyond the single-current level.

\section{Exact single-current dynamics in general non-Abelian WZW current-mode Lindbladians}
\label{Sec:Lindblad_general}

From the Lindbladian $SU(2)_k$ WZW CFTs, we have seen that symmetry of the jump rates under $q\to -q$ removes the bilinear terms in the Heisenberg equation of a single current mode and reduces the problem to a finite linear system in the Lie-algebra indices. 
We now show how this mechanism extends to a general non-Abelian WZW model based on a simple Lie algebra $\mathfrak g$.

As before, the coherent part is already closed:
\begin{equation}
	i[H_{\rm WZW},J_p^b]
	=
	i\frac{2\pi v}{L}[L_0,J_p^b]
	=
	-\,i\frac{2\pi v}{L}\,p\,J_p^b .
\end{equation}
For the dissipator, the general Heisenberg-picture formula is
\begin{equation}\label{eq:D_single_current_general_nonabelian}
	\mathcal{D}_{\rm WZW}^\dagger[J_p^b]
	=
	\sum_{a=1}^{\dim\mathfrak g}\sum_{q\in\mathbb Z}\frac{\gamma(q,a)}{2}
	\left(
	J_{-q}^a[J_p^b,J_q^a]
	+
	[J_{-q}^a,J_p^b]J_q^a
	\right).
\end{equation}
For generic rates $\gamma(q,a)$, this expression contains bilinear products of current modes and the single-current equation is not closed.

Similar to the $SU(2)_k$ case,
now let us impose symmetry under $q\to -q$ for each Lie-algebra index,
\begin{equation}
	\gamma(-q,a)=\gamma(q,a),
\end{equation}
and define
\begin{equation}
	\Gamma(a)=\frac{1}{2}\gamma(0,a)+\sum_{q>0}\gamma(q,a).
\end{equation}
Exactly as in the $SU(2)_k$ case, the bilinear terms then combine pairwise into commutators, and one finds
\begin{equation}
	\mathcal{D}_{\rm WZW}^\dagger[J_p^b]
	=
	\sum_{d=1}^{\dim\mathfrak g} M^{bd}J_p^d ,
\end{equation}
with
\begin{equation}
	M^{bd}
	=
	\sum_{a,c=1}^{\dim\mathfrak g}
	f^{abc}f^{acd}\, \Gamma(a).
\end{equation}
Thus the universal consequence of $q$-symmetric rates is that the Heisenberg equations for the family of single current modes $\{J_p^b\}$ close among themselves. In a general simple Lie algebra, however, the matrix $M^{bd}$ need not be diagonal in the chosen basis, so this gives in general a closed linear system rather than an equation closed on each individual $J_p^b$.

A further simplification occurs when the rates are also independent of the Lie-algebra index, i.e.,
\begin{equation}
	\gamma(q,a)=\gamma(q)=\gamma(-q).
\end{equation}
Then $\Gamma(a)\equiv \Gamma$ is independent of $a$, and the quadratic Casimir identity gives
\begin{equation}
	\sum_{a,c=1}^{\dim\mathfrak g}f^{abc}f^{acd}
	=
	-\,h^\vee\delta^{bd},
\end{equation}
where $h^\vee$ is the dual Coxeter number of $\mathfrak g$. Therefore,
\begin{equation}
	\mathcal{D}_{\rm WZW}^\dagger[J_p^b]
	=
	-\,h^\vee \Gamma\,J_p^b
	=
	-\frac{h^\vee}{2}\left(\sum_{q\in\mathbb Z}\gamma(q)\right)J_p^b ,
\end{equation}
and the full equation of motion closes on each individual current mode:
\begin{equation}
\label{eq:single_operator_EoM_nonAbelian}
	\frac{dJ_p^b}{dt}
	=
	-\left[
	i\frac{2\pi v}{L}p
	+
	\frac{h^\vee}{2}\sum_{q\in\mathbb Z}\gamma(q)
	\right]J_p^b .
\end{equation}

This is the general non-Abelian analogue of the single-current result. 
For the $SU(2)_k$ WZW theory, the dual Coxeter number is $h^\vee=2$, and Eq.~\eqref{eq:single_operator_EoM_nonAbelian} reduces to Eq.~\eqref{eq:single_operator_EoM_su2}. By contrast, for the Abelian $U(1)_k$ theory one has $h^\vee=0$, so the dissipative contribution vanishes identically, consistent with the discussion in Sec.\ref{Sec:Abelian_non-Abelian}.
Note that for a unitary WZW model based on a compact and non-Abelian simple Lie algebra $\mathfrak g$, the dual Coxeter number is always strictly positive, i.e., 
$h^\vee>0$. Therefore, the symmetric-rate single-current dynamics always exhibits exponential damping. The qualitative distinction between the Abelian and non-Abelian cases discussed in Sec.~\ref{Sec:Abelian_non-Abelian} is consequently not peculiar to the $SU(2)_k$ example, but follows generically from the positivity of $h^\vee$ in non-Abelian WZW current-mode Lindbladians.

As a remark, for arbitrary simple $\mathfrak g$, symmetric rates in $q$ are enough to reduce the dynamics of the current modes to a closed finite-dimensional linear system in the Lie-algebra indices. The stronger statement that each current mode evolves independently follows when the rates are also independent of the Lie-algebra index. The $SU(2)_k$ case discussed in the previous section is slightly special: there, the resulting linear system is already diagonal in the Cartesian basis, even before imposing index-independence.

\section{Discussion and conclusions}
\label{Sec:Discuss}

In this work, we introduced \textit{WZW current-mode Lindbladians}, namely Lindblad generators whose coherent part is given by the WZW Hamiltonian and whose jump operators are linear in affine current modes. Since the WZW Hamiltonian is quadratic in the currents through the Sugawara construction, the resulting Lindbladian is quadratic in the left- and right-multiplication superoperators associated with the current modes. At the same time, in the non-Abelian case this quadratic structure is not of the canonical quasi-free type familiar from third quantization, because the current modes satisfy an affine Kac--Moody algebra that cannot be reduced to canonical commutation or anticommutation relations.

For this reason, the exact solvability studied here is different from the usual quasi-free notion based on the full spectrum of the Lindbladian on Liouville space. Instead, we formulated exact solvability in terms of Heisenberg operator dynamics: the question is whether the equations of motion of current modes, or of composite operators built from them, close under the Lindbladian evolution. Whenever this happens, the time evolution can be determined exactly from the current algebra itself.

Within this framework, the Abelian and non-Abelian theories behave very differently. In the Abelian $U(1)_k$ case, the current algebra reduces to the Heisenberg algebra, and closure persists for arbitrary linear choices of current-mode jump operators. As a result, one obtains exact operator dynamics not only for single current modes but also for a broad class of composite operators built from several current modes, without imposing any restriction on the initial state. This makes the $U(1)_k$ theory a particularly transparent realization of the present construction. As an application, we showed that suitable asymmetric jump rates generate dissipative cooling toward the critical ground state of the $U(1)_k$ WZW theory, providing an exactly tractable critical-state preparation protocol.

The non-Abelian case is substantially more restrictive. For $SU(2)_k$, the structure-constant term in the affine algebra generates additional products of current modes in the dissipative evolution. 
At the level of a single current mode, symmetry of the jump rates under $q\to -q$, where $q$ denotes the mode index of the jump operators, is sufficient to reduce the equation of motion to a closed linear form. However, unlike in the Abelian
$U(1)_k$ theory, where the same symmetry condition eliminates the dissipative contribution entirely and leaves the single-current dynamics purely coherent, the non-Abelian structure constants produce an additional dissipative term. Consequently, the single-current equation remains closed but generally exhibits exponential damping.
 Furthermore, once products of two current modes are considered, the evolution generically couples them to infinitely many shifted products, so closure is lost even under the most symmetric choice of rates. This already demonstrates that the exact solvability found in the Abelian theory does not extend in any straightforward way to the non-Abelian setting.

We then generalized the single-current analysis to WZW models based on an arbitrary simple Lie algebra $\mathfrak g$. In general, symmetry of the jump rates $\gamma(q)$ under $q\to -q$ removes the quadratic terms in the Heisenberg equation of a single current mode and reduces the dynamics to a closed finite-dimensional linear system in the Lie-algebra indices. If the jump rates are furthermore independent of the Lie-algebra index, this system becomes diagonal and each current mode evolves independently with a decay rate fixed by the dual Coxeter number. Thus the $SU(2)_k$ result is not accidental, but reflects a general mechanism in non-Abelian WZW current-mode Lindbladians.

The above results lead to a clear overall picture: WZW current-mode Lindbladians provide a natural extension of exactly tractable Lindblad dynamics from the quasi-free setting to conformal field theories with affine current algebra. In the Abelian theory, this extension is broad and supports a wide class of exactly solvable dissipative dynamics. In the non-Abelian theory, by contrast, the same current algebra that makes the model interacting also strongly restricts the possibility of closed operator dynamics. In this precise sense, the symmetric infinite-temperature generator emerges as the distinguished non-Abelian case for which exact solvability survives, whereas no analogous exactly tractable non-Abelian cooling protocol is available within the present framework.

More broadly, the analysis shows that affine current algebra can serve as an organizing principle for exact Lindbladian dynamics in interacting conformal field theories, while also making clear where the obstruction appears in the non-Abelian case. It would be interesting to investigate whether related constructions exist for other classes of CFTs, for different choices of observables beyond products of current modes, or for more general dissipative deformations in which some remnant of the present algebraic solvability may survive.

Another interesting direction for future work is to test the dissipative cooling protocol introduced in Sec.~\ref{Sec:Cooling} in critical lattice systems. Dissipative approaches to preparing the ground states of many-body systems have attracted considerable recent attention \cite{2024_LinLin,2026_LinLin,2025_LinLin_Review}. However, to our knowledge, such protocols have not yet been systematically explored in quantum critical systems. We hope that the exactly solvable cooling scheme developed here in the $U(1)_k$ WZW CFT can provide useful insights into this problem and serve as a guiding framework for future studies on critical lattice models.

\smallskip
Finally, it would be interesting to extend the present analysis to time-dependent Lindbladians. Time dependence may be introduced either through the Hamiltonian or through the dissipative sector, potentially giving rise to novel nonequilibrium steady states and dynamical phases \cite{yoshida_2026}. An important question is whether the operator-closure structures identified in this work survive in such time-dependent settings, and if so, whether they can lead to new exactly solvable classes of driven open critical systems. 

One natural starting point is provided by the recently developed exactly solvable driven CFTs \cite{2018_Wen_Wu,2020_Fan,2020_Lapierre,
2021_Wen1,2020_Lapierre2,2022_Wen_random}, whose predictions have been observed experimentally in Ref.\cite{2026_Mo}. In these time-dependent driven CFTs, despite the explicit time dependence of the Hamiltonian, the operator dynamics can be analyzed analytically. It is therefore natural to ask how this picture is modified in the presence of dissipation. In particular, combining the driving protocols of driven CFTs with the cooling mechanism introduced in Sec.~\ref{Sec:Cooling} may lead to a rich interplay between coherent energy injection and dissipative energy removal. Since driven CFTs are known to exhibit both heating and non-heating phases \cite{2018_Wen_Wu}, one may expect qualitatively new dynamical regimes when dissipation is present. For example, in the heating phase, the competition between driving-induced heating and dissipative cooling could give rise to nontrivial nonequilibrium steady states, dynamical phase transitions, or other forms of universal critical behavior. Exploring these possibilities is left for a future study.


\begin{acknowledgments}
This work is supported by a startup at Georgia Institute of Technology. Q.T. thanks Naixu Guo and Zhong-Xia Shang for collaboration on a related project.
\end{acknowledgments}

\bibliography{ref}


\widetext
\appendix


\end{document}